\let\accentvec\vec 
\documentclass{llncs}
\let\vec\accentvec %

\usepackage[utf8]{inputenc}
\usepackage[T1]{fontenc}

\usepackage{todonotes}

\usepackage{amssymb}
\usepackage{amsmath}
\usepackage{amsfonts}

\usepackage[vlined,ruled]{algorithm2e}

\usepackage{graphicx}
\usepackage{wrapfig}
\usepackage{subfigure}
\usepackage{xcolor}

\usepackage{paralist}
\usepackage{hyperref}
\definecolor{darkblue}{rgb}{0,0,.5}
\hypersetup{colorlinks=true, breaklinks=true, linkcolor=darkblue, menucolor=darkblue, urlcolor=darkblue, citecolor=darkblue}

\newcommand{\gG}{G(n;c)}
\newcommand{\bG}{B(n/2,n/2;c)}

\newcommand{\Bin}{\mathrm{Bin}}
\newcommand{\Nor}{\mathrm{Nor}}
\newcommand{\abs}[1]{\lvert #1\rvert}
\newcommand{\figPath}{.}
\newcommand{\fr}{\lambda}
\newcommand{\mce}{\rho}
\newcommand{\mt}{\bar{t}}

\newcommand{\moos}{\overline{{\scriptstyle\#}o1}}
\newcommand{\mots}{\overline{{\scriptstyle\#}o2}}
\newcommand{\mhs}{\overline{{\scriptstyle\#}h}}

\spnewtheorem{fact}{Fact}{\bf}{\upshape}

\begin{document}
\title{A More Reliable Greedy Heuristic for Maximum Matchings in Sparse Random Graphs\texorpdfstring{\thanks{Research supported by DFG grant DI 412/10-2.}}{}}
\author{Martin Dietzfelbinger\inst{1} \and Hendrik Peilke\inst{2}\fnmsep\thanks{M.Sc. student at the Technische Universität Ilmenau while this work was done.} \and Michael Rink\inst{1}}

\institute{Fakultät für Informatik und Automatisierung, Technische Universität Ilmenau
\email{\{martin.dietzfelbinger,michael.rink\}@tu-ilmenau.de}%
\and%
IBYKUS AG für Informationstechnologie, Erfurt, Germany
\email{Hendrik.Peilke@Ibykus.de}%
}

\maketitle

\begin{abstract}
We propose a new greedy algorithm for the maximum cardinality matching problem. We give
experimental evidence that this algorithm is likely to find a maximum matching in random
graphs with constant expected degree $c>0$, independent of the value of $c$.
This is contrary to the behavior of commonly used greedy matching heuristics which are known
to have some range of $c$ where they probably fail to compute a maximum matching.
\end{abstract}
\section{Introduction}
\paragraph{Maximum Cardinality Matchings.}
Consider an undirected graph $G=(V,E)$ with node set $V$, $\abs{V}=n$, and edge set $E\subseteq \binom{V}{2}$, $\abs{E}=m$. 
A matching $M$ in $G$ is a subset of $E$
with the property that the edges in $M$ are pairwise disjoint.
The problem of finding a matching with the largest possible cardinality, a so called \emph{maximum matching},
has been a subject of study for decades. The first polynomial time algorithm for this problem
was given in 1965 by Edmonds~\cite{Edmonds:Paths:1965}. A straightforward
implementation of this algorithm has running time \mbox{$O(n^2\cdot m)$}.
Many other polynomial time algorithms followed,
eventually reducing the running time to $O(n^{1/2}\cdot m)$,
as, e.g., the algorithm of Micali and Vazirani~\cite{MV:Match:1980,Vazirani:Proof:1994}.
For dense graphs, i.e., graphs with $m=\Theta(n^2)$ edges, this was
the best known until 2004 when Mucha and Sankowski~\cite{MS:Match:2004} gave an
algorithm that has (expected) running time dominated by the time for
multiplying two $n\times n$ matrices, which is $O(n^\omega)$,
with $\omega<2.376$~\cite{CW:Matrix:1990}.
\paragraph{Heuristics.} 
Usually matching algorithms, notably augmenting path algorithms, are allowed to be initialized with a non-empty matching
which is then iteratively improved to a maximum matching.
Hence a large enough initial matching determined with some fast
heuristic approach can decrease the running time of an exact algorithm significantly.
Beyond the use of heuristics in the preprocessing phase of exact algorithms,
there is an interest in graph classes where heuristics, especially fast greedy algorithms, 
are likely to obtain maximum matchings. On such classes heuristics can replace the (overall) exact algorithms
if the heuristics are faster or at least equally fast but easier to implement.

\paragraph{Sparse Random Graphs.}
A well studied graph class in this context is the class of random graphs
with constant expected degree $c$.
Let $\gG$ be a random (general) graph with $n$ nodes where each of the $\binom{n}{2}$ possible edges is present with
probability $p=c/(n-1)$, and let $\bG$ be a random bipartite graph with $n$ nodes where each of the $n^2/4$ possible edges
is present with probability $p=c\cdot 2/n$.
Bast et al.~\cite{BMST:Match:2006} showed that if $c>c_0$  for $c_0=32.67$ in the case of general graphs, and $c_0=8.83$ in the case of bipartite graphs,
then with high probability every non-maximum matching in $\gG$ and $\bG$ has an augmenting path of length $O(\log n)$.
(Note that this trivially holds for $c\in(0,1)$ and indeed it is conjectured that $c_0=0$ in both cases.)
Hence matching algorithms using shortest augmenting paths like the algorithm of Micali and Vazirani for
general graphs and the algorithm of Hopcroft and Karp~\cite{HK:Match:1973} for bipartite graphs
have (expected) running time $O(n\cdot\log n)$ on sparse random graphs.
Chebolu et al.~\cite{CFM:Match:2010} gave an algorithm that improved the (expected) running time
to $O(n)$ using a simple heuristic in the first phase of their algorithm,
usually called Karp-Sipser algorithm. Karp and Sipser~\cite{KS:Match:1981} proved that 
this greedy algorithm produces a matching which is within $o(n)$
of the maximum for every constant $c>0$. This result was improved by Aronson et al.~\cite{AFP:Match:1998}
who showed that actually for $c<e$ the Karp-Sipser algorithm finds a maximum matching with high probability
and for $c>e$ the size of the matching is within $n^{1/5+o(1)}$ of the maximum.
Interestingly, for practical purposes Karp and Sipser suggested a different greedy algorithm, Algorithm 1 of~\cite{KS:Match:1981},
that turns out to give better results in their experiments but seems to be much more complicated to
analyze because it utilizes contraction of nodes.
\paragraph{``Critical Region''.}
In an experimental study Magun~\cite{Magun:Exp:1998} compared the performance of several greedy matching algorithms in
the style of the algorithms given in~\cite{KS:Match:1981} on sparse random graphs. It turned out that
there are good greedy algorithms that are likely to give maximum matchings for 
a wide range of $c$, but even the best algorithm in this study fails in the range of about $2.6\leq c\leq 3.8$
(where the lower bound is likely to converge to $e\approx 2.718$ for $n$ large enough).
Hence there is some region for $c$ that seems critical for known greedy matching heuristics.

\subsection{Our Results}
We describe a new greedy matching algorithm
and give experimental evidence that this algorithm
is likely to compute a maximum matching in sparse random
graphs for all ranges of $c$ and large enough $n$;
in particular, it seems to overcome
the critical region mentioned in~\cite{Magun:Exp:1998}.
The algorithm is motivated by the ``self\/less algorithm'' of Sanders~\cite{Sanders:Self:2004},
for orienting undirected graphs such that the maximum in-degree is below a given constant.
\paragraph{Drawback.} In comparison to the common greedy heuristics discussed above
the running time of our algorithm is larger and
more affected by the expected degree $c$.
Hence, we propose using a combined algorithm using our approach solely for the critical region.

\subsection{Overview of the Paper}
In the next section we consider several common greedy matching heuristics
and give some motivation for our new approach. 
Following that, in the main part of the paper we describe the experiments and discuss the results.

\section{Greedy Matching Heuristics}
In this section we give a brief description of the greedy matching algorithms considered here.
The structure of this section is similar to Section 3 of~\cite{Magun:Exp:1998}.

\paragraph{Basic Structure.}
The algorithms work recursively. 
Let $G_0=G$ be the input graph. Consider some arbitrary recursion level $l\geq0$.
Let $G_{l}$ be the current graph, and let $d$ be the minimum degree of $G_l$. There are two cases:
\begin{compactitem}
\item[$d\leq 2$.] Apply an ``optimal reduction step'' on $G_{l}$, i.e., 
depending on $d$, remove nodes and edges from $G_{l}$ to yield $G_{l+1}$.
\item[$d\geq 3$.] Apply a ``heuristic reduction step'' on $G_{l}$, i.e.,
choose an edge $e=\{u,v\}$ from $G_{l}$ with the highest priority
according to some heuristic order of priority, and remove $u$ and $v$ and all incident edges from $G_{l}$ to yield $G_{l+1}$.
\end{compactitem}
Run the algorithm recursively on $G_{l+1}$, which will return a matching $M_{l+1}$ for $G_{l+1}$. Finally,
add an edge to $M_{l+1}$ to obtain a matching $M_{l}$ for $G_{l}$.
An optimal step will never decrease the size of a maximum matching, while a heuristic step might do that.
\paragraph{Optimal Steps.}
The two optimal steps that we consider are commonly known as ``degree 1 reduction'' and ``degree 2 reduction''.
They are based on the following facts proved by Karp and Sipser in~\cite{KS:Match:1981}.
\begin{fact}
Let $G=(V,E)$ be a graph. If there exists a node $u\in V$ with degree $\deg(u)=1$,
adjacent to a node $v\in V$, then there exists a maximum matching $M$ in $G$ with
$\{u,v\} \in M$.
\end{fact}
\begin{fact}
Let $G=(V,E)$ be a graph. If there exists a node $u\in V$ with degree $\deg(u)=2$,
adjacent to nodes $v_1,v_2\in V$, then there exists a maximum matching $M$
in $G$ with either $\{u,v_1\}\in M$ or $\{u,v_2\}\in M$.
\end{fact}
For any subset $V'$ of the nodes of $G$ let $G\setminus V'$ be the subgraph of $G$
that is induced by all nodes of $V\setminus V$' and let $G \circ V'$ be
the graph that results from $G$ by contracting all nodes of $V'$ into a single node 
and removing all multiple edges and self-loops.
Using these definitions we can state the optimal degree reduction steps as follows.
\begin{compactdesc}
\item[degree 1 reduction:] Randomly choose a node $u$ from $G_l$ with degree $\deg(u)=1$,
incident to an edge $e$. Shrink the graph $G_l$ via $G_{l+1}\leftarrow G_l \setminus e$.
Increase the matching $M_{l+1}$ given by the recursive call, via $M_l \leftarrow M_{l+1}\cup \{e\}$. 
\item[degree 2 reduction:] Randomly choose a node $u$ from $G_l$ with degree $\deg(u)=2$,
 adjacent to nodes $v_1,v_2$. Contract the three nodes into a single node $v$
via $G_{l+1}\leftarrow G_l\circ\{u,v_1,v_2\}$ and store how $v$ was constructed.
If an edge $e=\{v,w\}$ is part of the matching $M_{l+1}$ given by the recursive call,
then, to obtain the matching $M_l$, either replace $e$ with $\{v_1,w\}$ in $M_{l+1}$ and add $\{u,v_2\}$ to $M_{l+1}$,
or replace $e$ with $\{v_2,w\}$ in $M_{l+1}$ and add $\{u,v_1\}$ to $M_{l+1}$.
\end{compactdesc}
In the following we will use ``OPT(1)'' and ``optimal degree 1 reduction'',
as well as ``OPT(1,2)'' and ``optimal degree 1 and optimal degree 2 reduction''
synonymously.

\paragraph{Heuristic Steps.}
The procedure of the heuristic step is similar to the degree 1 reduction step.
First choose an edge $e$, then shrink the graph via $G_{l+1}\leftarrow G_l\setminus e$,
and finally increase the matching via $M_l \leftarrow M_{l+1}\cup \{e\}$.
The choice of the edge is based on a priority order of the edges,
where the priorities are calculated using
properties in the neighborhood of the nodes.
We consider the following heuristics.
\begin{compactdesc}
\item[random edge:] Randomly choose an edge $e\in E$.
\item[double minimum degree:] Randomly choose a node $u\in V$ among the nodes with smallest degree.
Randomly choose an edge $e=\{u,v\}\in E$ where $v$ is among the neighbors of $u$ that have smallest degree.
\item[minimum expected potential, minimum degree:]\hspace{-4.6pt} Randomly choose a node $u\in V$ among the nodes with smallest potential $\pi(u)$, where
\begin{displaymath}
\pi(u)=\sum\nolimits_{\{u,v\}\in E} \frac{1}{\deg(v)} \ . 
\end{displaymath}
Then randomly choose an edge $e=\{u,v\}\in E$ where $v$ is among the neighbors of $u$ that have smallest degree.
\end{compactdesc}
Simply choosing an edge at random can be seen as all edges having the same
priority, which disregards the structure of the graph.
The idea of choosing a node of low degree is that
the lower the degree the fewer the possibilities of the node $u$ to be covered by a matching.
This is taken one step further in the third heuristic by calculating the
values $\pi(u)$. If each neighbor $v$ of a node $u$ randomly declares one of its incident edges to be
the only edge that is allowed to cover $v$ in a matching then the value $\pi(u)$
is the expected number of potential matching edges that could cover $u$.
As before, the lower the number of possibilities the more urgent it is to
include the node in a matching edge.

In the following we will use interchangeably:
``HEU(rand)'' and ``random edge heuristic'',
``HEU(deg,deg)'' and ``double minimum degree heuristic'', as well as,
``HEU(pot,deg)'' and ``minimum expected potential, minimum degree heuristic''.

\paragraph{Algorithms.}
\label{sec:algorithms}
We list six matching algorithms whose performance is experimentally examined in 
our experiments, where the last two algorithms are new. 
The names of the algorithms are generic, describing their structure
as combination of the utilized optimal and heuristic steps.
If an algorithm uses OPT(1,2) then the degree 1 reduction step is always preferred to
the degree 2 reduction step.
\begin{compactdesc}
 \item[OPT(1):HEU(rand)]
This algorithm is commonly known as Karp-Sipser algorithm
as it was first analyzed by Karp and Sipser in~\cite[Algorithm 2]{KS:Match:1981}.
If the expected degree $c$ of a sparse random graph is below $e$ then the
algorithm finds a maximum matching (with high probability) and if
$c$ is larger than $e$ then the matching is within $n^{1/5+o(1)}$ of the maximum cardinality (with high probability), see~\cite{AFP:Match:1998}.

\item[OPT(1,2):HEU(rand)]
This is a variant of the Karp-Sipser algorithm using in addition
the degree 2 reduction step, which was also proposed in~\cite{KS:Match:1981}.
It is included to investigate the effect of the degree 2 reduction.

\item[OPT(1):HEU(deg,deg)]
This algorithm is recommended in the experimental study~\cite{Magun:Exp:1998} as the most practical algorithm, see~\cite[Conclusion]{Magun:Exp:1998}. 
Note that the optimal degree 1 reduction needs not to be implemented separately since
it is performed implicitly by the heuristic step.

\item[OPT(1,2):HEU(deg,deg)]
This is one of the two algorithms proposed in~\cite{Magun:Exp:1998} that offer the highest quality of solution.
The other one (called BlockRed) is more complicated,
using an additional optimal reduction, but has very similar performance.
It was demonstrated experimentally that both algorithms are likely to compute a maximum matching in sparse random graphs
when $c< 2.6$ or $c>3.8$, but fail to do so for other values of $c$.
Moreover, in the ``critical region'' $2.6\leq c\leq 3.8$ the number of edges that are missing from a matching with maximum cardinality
is increasing with increasing $n$.

\item[OPT(1):HEU(pot,deg)]
This is the first new algorithm. It is a straightforward adaption of the self\/less algorithm proposed by Sanders in~\cite{Sanders:Self:2004}
for determining an orientation of the edges of an undirected graph. The self\/less algorithm
has been proven to be optimal in the sense that with high probability it obtains an orientation of the edges of an undirected sparse random graph
that gives minimum in-degree, if the density is such that such an orientation exists, see~\cite{CSW:Self:2007}.

\item[OPT(1,2):HEU(pot,deg)]
This is the second new algorithm and the outcome of our search for an algorithm that has probably no critical region.
As shown in the following experiments the additional use of the degree 2 reduction is essential.
\end{compactdesc}
Note that the recursive structure of the algorithms can easily be transformed into an iterative structure,
if there is no degree 2 reduction or one only needs to compute the size of a maximum matching,
since in both cases there is no need to resolve contraction of nodes.

Algorithm OPT(1,2):HEU(pot,deg) is the heuristic that we propose
for computing maximum cardinality matchings in sparse random graphs, therefore its pseudocode (Algorithm~\ref{alg:opt}) is given below for completeness.

\newcommand{\algo}[1]{ \texttt{OPT(1,2):HEU(pot,deg)[}#1\texttt{]}}
\begin{algorithm}
\caption{\texttt{OPT(1,2):HEU(pot,deg)[$G$: graph]}}\label{alg:opt}
 \SetKw{KwInput}{Input:}
 \SetKw{KwOutput}{Output:}
 \SetNoFillComment

  \KwInput{simple graph $G=(V,E)$ with node set $V$ and edge set $E$}\\
  \KwOutput{matching $M$}\\
  \BlankLine
  $M\leftarrow \emptyset$\;
  \BlankLine
 
  \If{$E \not=\emptyset$}
  {

    $d \leftarrow $minimum degree of all nodes in $V$\;
    \If{ $d=1$}
    { 
      $u\leftarrow$ random node from $V$ with $\deg(u)=1$\;
      $v\leftarrow$ neighbor of $u$\;
      $M\leftarrow \algo{G\setminus \{u,v\}}$\;
      $M\leftarrow M\cup \{ \{u,v\}\}$;
    }

    \ElseIf{$d=2$}
    {
       $u\leftarrow$ random node from $V$ with $\deg(u)=2$\;
       $\{v_1,v_2\}\leftarrow$ set of 2 neighbors of $u$\;
       $v\leftarrow\{u,v_1,v_2\}$\;
       $M\leftarrow \algo{G\circ \{u,v_1,v_2\}}$\;
       \lIf{\upshape$v$ is not matched in $M$}
       {
         $M\leftarrow M\cup \{ \{u,v_1\}\}$\;
       }
       \Else
       {
        $w\leftarrow$ matching neighbor of $v$\;
        $M\leftarrow M \setminus \{ \{v,w\}\}$\;
        \lIf{$\{v_1,w\}\in E$}
        {
          $M\leftarrow M\cup \{ \{v_1,w\}, \{u,v_2\}\}$\;
        }
        \lElse
        {
          $M\leftarrow M\cup \{ \{v_2,w\}, \{u,v_1\}\}$\;
        }
      }
    }
    \Else
    {
     $\pi \leftarrow $minimum potential of all nodes in $V$\;
     $u\leftarrow$ random node from $V$ with $\pi(u)=\pi$\;
     $N\leftarrow$ set of neighbors of $u$\;
     $v\leftarrow$ random node from $N$ with minimum degree\;
     $M\leftarrow \algo{G\setminus \{u,v\}}$\;
     $M\leftarrow M\cup \{ \{u,v\}\}$;
    }
  }
\KwRet $M$\;
\end{algorithm}

\section{Experiments}
We examine the performance of the six greedy matching algorithms, given in the last section,
on random general graphs $\gG$ and random bipartite graphs $\bG$ 
with $n$ nodes and constant expected average degree $c$.
We cover parameter ranges $n\in\{10^4,10^5,10^6\}$ and $c\in[1,10]$, where parameter
$c$ is iteratively increased via $c=1+i\cdot 0.1$, for $i=0,1,\ldots,90$.
\paragraph{Construction of Random Graphs.}
Let $N=\binom{n}{2}, p=c/(n-1)$ for random general graphs $\gG$,
and let $N=n^2/4, p=c\cdot 2/n$ for random bipartite  graphs $\bG$.
For fixed parameters $(n,c)$ the construction of a random graph $G=(V,E)$ is done as follows.
We start with the node set $V=\{1,2,\ldots,n\}$ and an empty edge set $E$.
If $n=10^4$ then each of the $N$ possible edges is generated and added to $E$ with
probability $p$ independently of all other edges.
If $n\in\{10^5,10^6\}$ then, in order to keep the construction time manageable, we
first determine the number of edges $X$, which is expected to be linear in $n$, and then randomly choose $X$ edges from the
set of $N$ possible edges. The number of edges follows a binomial distribution $X\sim\Bin(N,p)$.
To determine a realization $x$ of $X$, we determine a realization $y$ of a standard normal random variable $Y\sim\Nor(0,1)$ using
the polar method~\cite[Section 2.3.1]{TLLV:Gauss:2007}.
The value  $\tilde{x}= \mathrm{round}(y \cdot \sqrt{ N\cdot p (1-p)} + N\cdot p)$ is used as an approximation of $x$.
As long as $\tilde{x}$ is not feasible the calculation is repeated with new realizations of $N$.
\paragraph{Measurements.}
For each pair of parameters $(n,c)$ we constructed $100$ random graphs (bipartite and general) and
measured the following quantities for each of the six heuristics:
\begin{compactitem}
 \item the failure rate $\fr$. This is the fraction of graphs where the matching obtained by the heuristic is not a maximum matching.
 \item the average number of ``lost edges'' $\mce$, which we define as the average number of edges missing
       from a maximum matching, conditioned on the event that a failure occurs. If no failure occurs we let $\mce=0$.
\end{compactitem}
To get insight in how the parameter $c$ might influence the running time of our new algorithm we did 
additional experiments using random graphs with $n=10^6$ nodes. 
For each $c$ we constructed $10$ random graphs (bipartite and general) and
measured the following quantities for OPT(1,2):HEU(pot,deg):
\begin{compactitem}
 \item the average running time $\mt$ needed to obtain a matching, as well as
the corresponding sample variance.
 \item the average fraction of: degree 1 reduction steps $\moos$, degree 2 reduction steps $\mots$, and
heuristic steps $\mhs$.
\end{compactitem}

\paragraph{System.} 
The source code for the graph generators as well as for the algorithms is written in C++ and compiled with
g++ version 4.5.1. The experiments regarding the running time ran on an Intel Xeon CPU E5450 (using one core)
under openSUSE with kernel 2.6.37.6-0.9-desktop.
\paragraph{Random Source.}
For the necessary random choices for the algorithms as well as for the construction
of the random graphs we used the pseudo random number generator {$\mathrm{MT}19937$} ``Mersenne Twister''
of the GNU Scientific Library~\cite{GNU:Scientific:2011}.
\subsection{Results}
Here we consider results from the matching heuristics given in Section~\ref{sec:algorithms}.

\begin{figure}[htb]
\vspace{-2cm}
\setlength{\subfigcapskip}{-0.5cm}
\centering
\hspace{-0cm}
\vspace{-1.5cm}
\subfigure[general graphs]{\scalebox{0.75}{\input{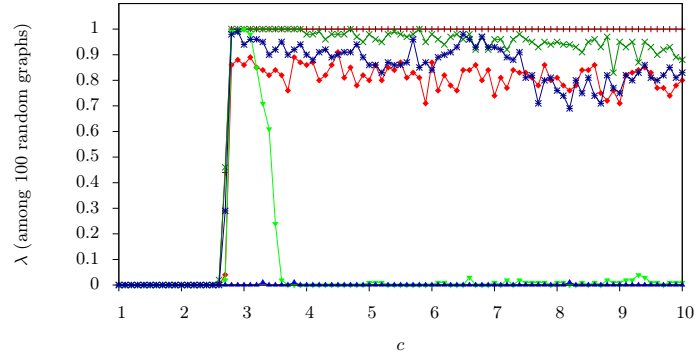}}}
\subfigure[bipartite graphs]{\scalebox{0.75}{\input{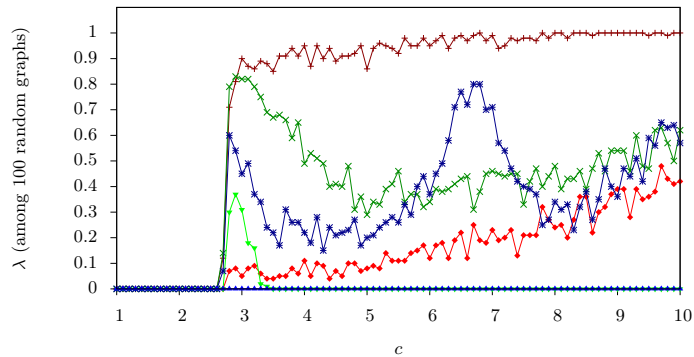}}}
\vspace{-0.3cm}
\caption{Failure rates on graphs with $n=10^6$ nodes.}
\vspace{-0.5cm}
\label{fig:failure_rate_n=10^6}
\end{figure}

\paragraph{Failure Rates.}
\begin{wrapfigure}{r}{5cm}
\vspace{-0.8cm}
  \includegraphics[width=5cm]{\figPath/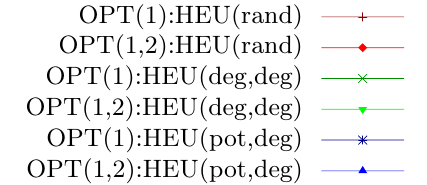}
\vspace{-1cm}
\end{wrapfigure}
Figure~\ref{fig:failure_rate_n=10^6} gives the failure rates on 
general and bipartite random graphs with $n=10^6$ nodes and expected degree $c$ ranging from $1$ to $10$. 
The legend for both plots is given to the right.
Figures depicting the failure rates for graphs with $10^4$ and $10^5$ nodes are given in Appendix~\ref{app:failure_rate}.
The results are qualitatively similar to the results for $n=10^6$.

For $1\leq c \leq 2.5$ no failure occurred in any of the algorithms.
Our new algorithm OPT(1,2):HEU(pot,deg) never failed on bipartite graphs and failed three times
on general graphs, for $c\in\{3.3,3.8,8.2\}$ with failure rate $\fr=1/100$. For the other algorithms we observed the
following behavior.
\begin{compactitem}
 \item 
For general graphs at $c=2.8$ all of them have a failure rate $\fr$ of at least $0.86$.
For OPT(1,2):HEU(deg,deg) we could replicate the behavior, observed in~\cite{Magun:Exp:1998}, that
for $c\leq 2.6$ and $c\geq 3.7$ the failure rate of the algorithm is almost zero while for the other values of $c$ the failure
rate is very high, reaching its peak with $\fr=1$ at $c=3.0$.
For the other heuristics $\fr$ stays quite high after $c=2.8$.
\item For bipartite graphs the situation is different. The failure rates go up only beyond $2.6$
and the qualitative behavior varies widely among the different heuristics.
For OPT(1,2):HEU(deg,deg) we observed a critical region of $2.8<c<3.5$ but 
with a less pronounced failure rate, reaching its peak at $c=2.9$ with $\fr=0.37$.
For all other algorithms the failure rate seems to increase for $c$ beyond $8$.
\end{compactitem}
It is proven that OPT(1):HEU(rand) is likely to find a maximum matching for $c<e\approx 2.718$
mainly due to the optimal degree 1 reduction steps (so called $e$-phenomenon), see~\cite{AFP:Match:1998}. Our results indicate that including degree 2 reductions does not influence this bound much.
Overall, the heuristics with degree 2 reduction more often give a maximum matching than their counterparts that can only utilize degree 1 reduction.
In terms of the difference of the failure rates this effect is smallest for OPT(1):HEU(rand) and OPT(1,2):HEU(rand) on general random graphs.
The best algorithms in terms of quality of solution are OPT(1,2):HEU(deg,deg) and OPT(1,2):HEU(pot,deg).

\begin{figure}[ht]
\vspace{-2cm}
\setlength{\subfigcapskip}{-0.5cm}
\centering
\hspace{-0cm}
\vspace{-1.5cm}
\subfigure[general graphs]{\scalebox{0.75}{\input{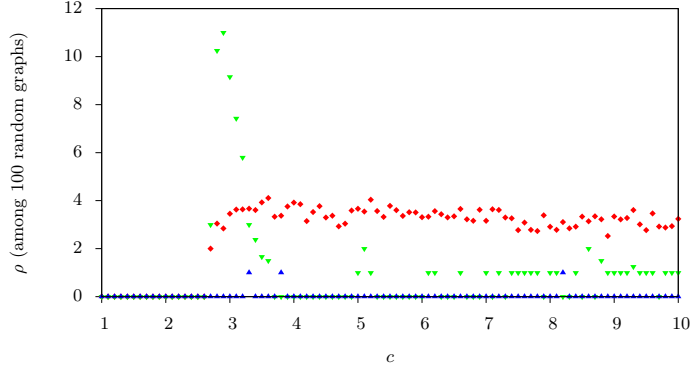}}}
\subfigure[bipartite graphs]{\scalebox{0.75}{\input{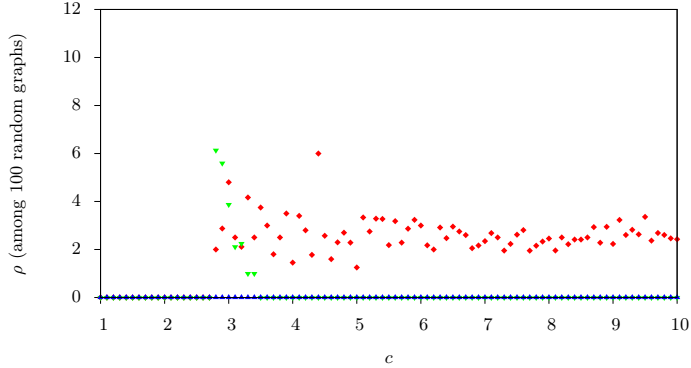}}}
\vspace{-0.3cm}
\caption{Average number of lost edges (if $\fr>0$) for graphs with $n=10^6$ nodes.}
\vspace{-0.5cm}
\label{fig:lost_edges_n=10^6}
\end{figure}

\paragraph{Edges Lost if Failure Occurs.}
\begin{wrapfigure}{r}{5cm}
\vspace{-1cm}
    \includegraphics[width=5cm]{\figPath/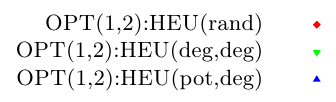}
\vspace{-1cm}
\end{wrapfigure}
Unlike before, we are only interested in the algorithms using degree 2 reduction, since on average they obtain the largest matchings.
Figure~\ref{fig:lost_edges_n=10^6} gives the average number of lost edges conditioned on the event that a failure occurs,
for general and bipartite random graphs with $n=10^6$ nodes and expected degree $c$ ranging from $1$ to $10$.
The legend for both plots is given on the top right of this paragraph.
The figures for the number of lost edges for graphs with $10^4$ and $10^5$ nodes are given in Appendix~\ref{app:lost_edges}.
The results are qualitatively similar to the results for $n=10^6$.

The mean over the values $\mce$ for heuristic OPT(1,2):HEU(rand) is higher for the general graph scenario than for
the bipartite graph scenario, while the variance of $\mce$ is lower.
The number of lost edges for heuristic OPT(1,2):HEU(rand) and for heuristic OPT(1,2):HEU(deg,deg), within their critical ranges,
increases with increasing $n$, cf.~Appendix~\ref{app:lost_edges}. 
Outside its critical range the double minimum degree heuristic OPT(1,2):HEU(deg,deg) loses mostly one edge on average for fixed $c$ on general graphs
and no edge on bipartite graphs. Our new algorithm OPT(1,2):HEU(pot,deg) loses one edge only in three cases.

\begin{figure}[htb]
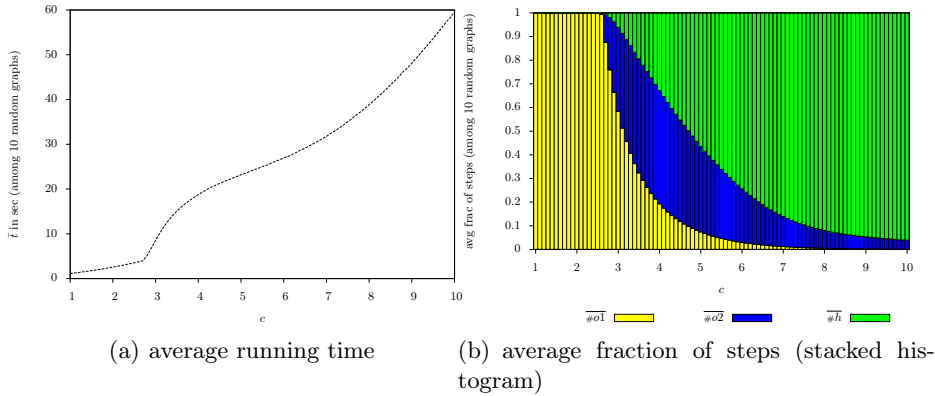

\hspace{-0.25cm}
\subfigure[average running time]{\scalebox{0.5}{\input{\figPath/new_random_general_graphs__nodes=10E6_attempts=10E1_degrees=1-10_v1__runningTime_TEX.tex}}}
\hspace{-0.45cm}
\subfigure[average fraction of steps (stacked histogram)]{\scalebox{0.5}{\input{\figPath/new_random_general_graphs__nodes=10E6_attempts=10E1_degrees=1-10_v1__fracOfSteps_TEX.tex}}}
\caption{Run-time behavior of algorithm OPT(1,2):HEU(pot,deg) on general random graphs with $10^6$ nodes.}
\label{fig:run-time_behavior}
\end{figure}

\paragraph{Run-time Behavior.}
Figure~\ref{fig:run-time_behavior} shows the average running time $\mt$ of algorithm OPT(1,2):HEU(pot,deg) 
for calculating a matching, as well as the corresponding average fraction of degree~1 reduction steps $\moos$, degree~2 reduction steps $\mots$, and heuristic steps $\mhs$,
on general random graphs with $10^6$ nodes.
The run-time behavior on bipartite random graphs of this size is qualitatively and quantitatively quite similar and given
in Appendix~\ref{app:run-time_behavior}. The failure rate was zero in these experiments.

The average running time exhibits a non-linear increase.
In a first phase, for $1\leq c\leq 2.8$, the slope is linear and quite low.
This is because in this range the running time is dominated by the fraction of degree 1 reduction steps $\moos$ which is more than 99 percent.
It follows a second phase starting with a sudden increase of $\mt$ which starts to flatten soon at $c$ about $3.5$.
This goes along with a strong decrease of $\moos$ and increase of $\mots$ and $\mhs$.
The next slight increase of the slope seems to be between $c=6$ and $c=7$ 
when $\moos$ falls below $0.03$ and the fraction of heuristic steps $\mhs$ is more than $0.7$,
which indicates the begin of a third phase.
The slope in this phase is larger than in the first phase and seems to be slightly non-linear.
The sample variance of the running time is very low for the first phase and then increases slightly with increasing $c$;
we observed a maximum of about $0.29$ for general random graphs and of about $0.38$ for bipartite random graphs.

\section{Summary and Future Work}
We proposed a new greedy algorithm to solve the maximum cardinality matching problem on random graphs with constant
expected degree $c$, and found in experiments that this algorithm has a very low failure rate for a broad range of~$c$.
It is an open problem to prove that this behavior is to be expected.

The algorithm itself is an adaption of the self\/less algorithm of Sanders~\cite{Sanders:Self:2004} for orienting graphs,
which was successfully generalized to orienting hypergraphs before, see~\cite{DGMMPR:XORSAT:2010}.
It seems possible that the ``self\/less approach'' can be used as generic building block for other
greedy algorithms on random graphs too, like, e.g., graph coloring, which 
would be interesting to investigate.

\vfill
\pagebreak
\appendix

\newcommand{\imgScale}{0.75}
\section{Failure Rates}
\label{app:failure_rate}
\begin{figure}[htb]
\vspace{-2.2cm}
\setlength{\subfigcapskip}{-1.5cm}
\centering
\hspace{-0cm}
\vspace{-2.2cm}
\subfigure{\scalebox{\imgScale}{\input{\figPath/random_general_graphs__nodes=10E4_attempts=10E2_degrees=1-10__relError_TEX.tex}}}
\vspace{-2.2cm}
\subfigure{\scalebox{\imgScale}{\input{\figPath/random_general_graphs__nodes=10E5_attempts=10E2_degrees=1-10__relError_TEX.tex}}}
\subfigure{\scalebox{\imgScale}{\input{\figPath/random_general_graphs__nodes=10E6_attempts=10E2_degrees=1-10__relError_TEX.tex}}}
\vspace{-2.2cm}
\includegraphics[height=2cm]{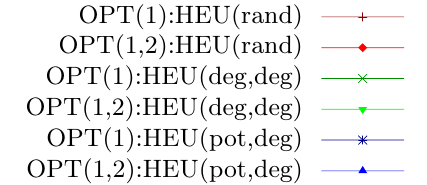}
\vspace{-1cm}
\caption{General random graphs. Number of nodes: 1st $n=10^4$, 2nd $n=10^5$, 3rd $n=10^6$.}
\end{figure}

\begin{figure}[htb]
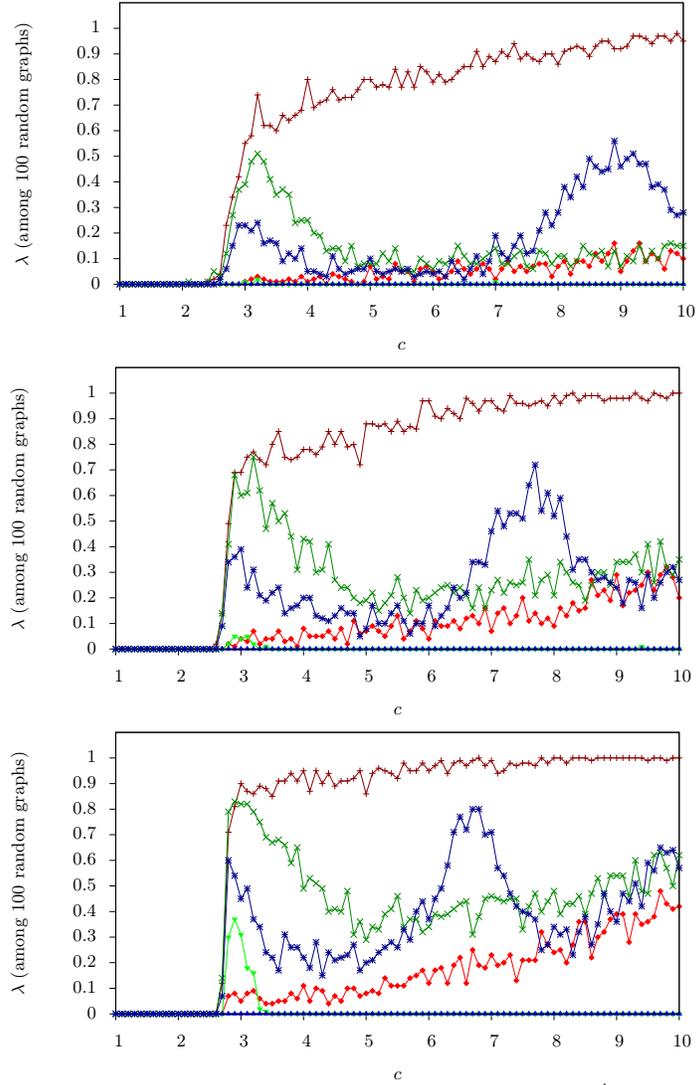

\vspace{-2.2cm}
\setlength{\subfigcapskip}{-1.5cm}
\centering
\hspace{-0cm}
\vspace{-2.2cm}
\subfigure{\scalebox{\imgScale}{\input{\figPath/random_bipartite_graphs__nodes=10E4_attempts=10E2_degrees=1-10__relError_TEX.tex}}}
\vspace{-2.2cm}
\subfigure{\scalebox{\imgScale}{\input{\figPath/random_bipartite_graphs__nodes=10E5_attempts=10E2_degrees=1-10__relError_TEX.tex}}}
\subfigure{\scalebox{\imgScale}{\input{\figPath/random_bipartite_graphs__nodes=10E6_attempts=10E2_degrees=1-10__relError_TEX.tex}}}
\vspace{-2.2cm}
\includegraphics[height=2cm]{\figPath/legend_failure_rate.png}
\vspace{-1cm}
\caption{Bipartite random graphs. Number of nodes: 1st $n=10^4$, 2nd $n=10^5$, 3rd $n=10^6$.}
\end{figure}

\clearpage
\section{Average Number of Lost Edges if Failure Occurs}
\label{app:lost_edges}
\begin{figure}[htb]
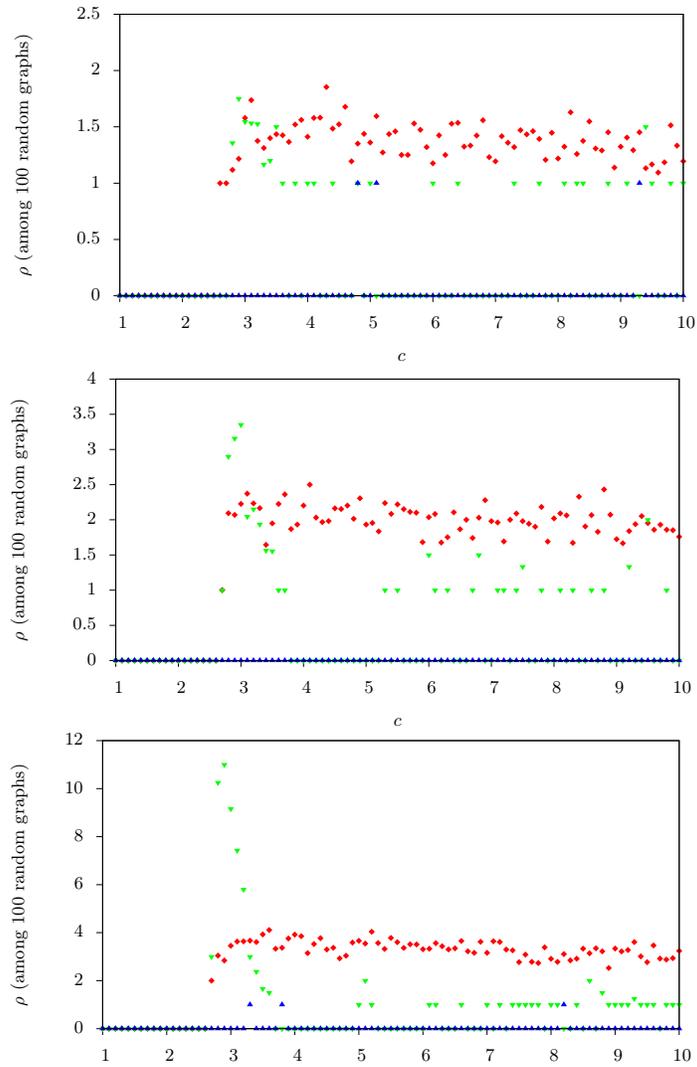

\vspace{-2.2cm}
\setlength{\subfigcapskip}{-1.5cm}
\centering
\hspace{-0cm}
\vspace{-2.2cm}
\subfigure{\scalebox{\imgScale}{\input{\figPath/random_general_graphs__nodes=10E4_attempts=10E2_degrees=1-10__absError_TEX.tex}}}
\vspace{-2.2cm}
\subfigure{\scalebox{\imgScale}{\input{\figPath/random_general_graphs__nodes=10E5_attempts=10E2_degrees=1-10__absError_TEX.tex}}}
\subfigure{\scalebox{\imgScale}{\input{\figPath/random_general_graphs__nodes=10E6_attempts=10E2_degrees=1-10__absError_TEX.tex}}}
\vspace{-1.5cm}
\includegraphics[height=1.3cm]{\figPath/legend_lost_edges.png}
\vspace{-1cm}
\caption{General random graphs. Number of nodes: 1st $n=10^4$, 2nd $n=10^5$, 3rd $n=10^6$.}
\end{figure}

\begin{figure}[htb]
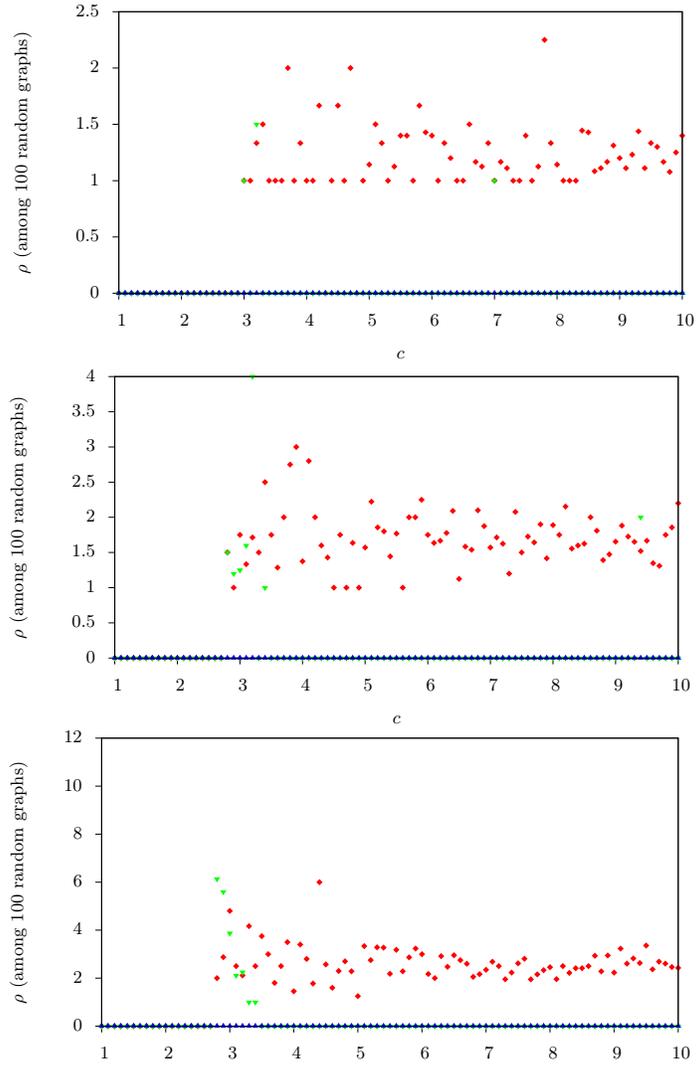

\vspace{-2.2cm}
\setlength{\subfigcapskip}{-1.5cm}
\centering
\hspace{-0cm}
\vspace{-2.2cm}
\subfigure{\scalebox{\imgScale}{\input{\figPath/random_bipartite_graphs__nodes=10E4_attempts=10E2_degrees=1-10__absError_TEX.tex}}}
\vspace{-2.2cm}
\subfigure{\scalebox{\imgScale}{\input{\figPath/random_bipartite_graphs__nodes=10E5_attempts=10E2_degrees=1-10__absError_TEX.tex}}}
\subfigure{\scalebox{\imgScale}{\input{\figPath/random_bipartite_graphs__nodes=10E6_attempts=10E2_degrees=1-10__absError_TEX.tex}}}
\vspace{-1.5cm}
\includegraphics[height=1.3cm]{\figPath/legend_lost_edges.png}
\vspace{-1cm}
\caption{Bipartite random graphs. Number of nodes: 1st $n=10^4$, 2nd $n=10^5$, 3rd $n=10^6$.}
\end{figure}
\clearpage
\section{Average Running Times and Average Fraction of Steps}
\label{app:run-time_behavior}
\begin{figure}[htb]
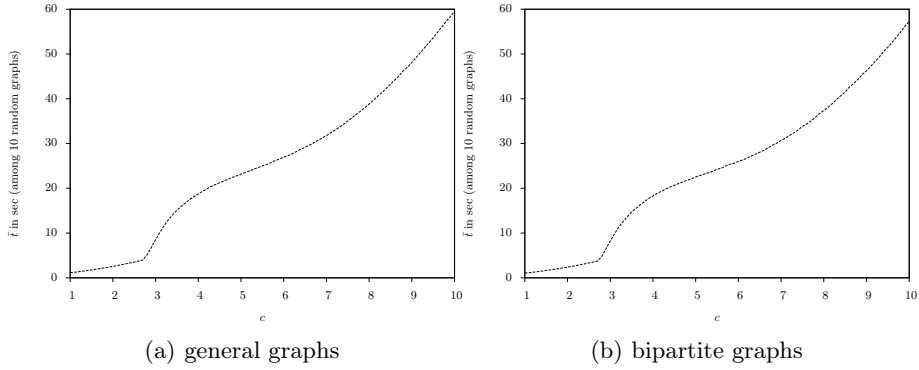

\vspace{-1cm}
\hspace{-0.25cm}
\subfigure[general graphs]{\scalebox{0.5}{\input{\figPath/new_random_general_graphs__nodes=10E6_attempts=10E1_degrees=1-10_v1__runningTime_TEX.tex}}}
\hspace{-0.45cm}
\subfigure[bipartite graphs]{\scalebox{0.5}{\input{\figPath/new_random_bipartite_graphs__nodes=10E6_attempts=10E1_degrees=1-10_v1__runningTime_TEX.tex}}}
\caption{Average running times in seconds for OPT(1,2):HEU(pot,deg) to obtain a matching on random graphs with $n=10^6$ nodes.}
\end{figure}

\begin{figure}[htb]
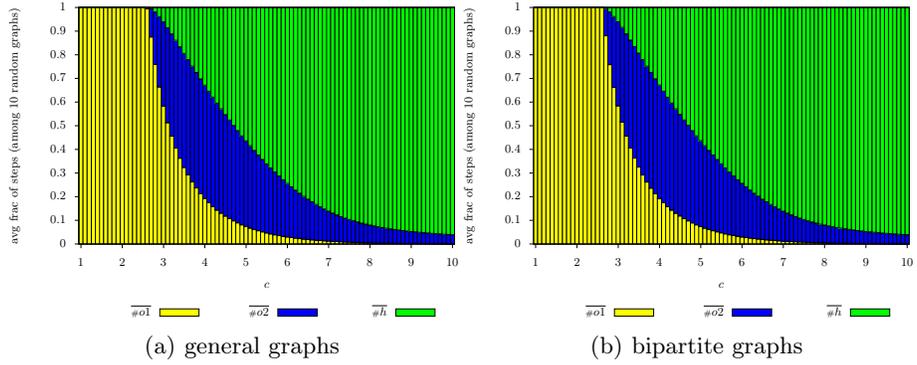

\hspace{-0.25cm}
\subfigure[general graphs]{\scalebox{0.5}{\input{\figPath/new_random_general_graphs__nodes=10E6_attempts=10E1_degrees=1-10_v1__fracOfSteps_TEX.tex}}}
\hspace{-0.45cm}
\subfigure[bipartite graphs]{\scalebox{0.5}{\input{\figPath/new_random_bipartite_graphs__nodes=10E6_attempts=10E1_degrees=1-10_v1__fracOfSteps_TEX.tex}}}
\caption{Stacked histogram of average fraction of degree 1 reduction steps, degree 2 reduction steps, and heuristic steps, 
of algorithm OPT(1,2):HEU(pot,deg) on random graphs with $n=10^6$ nodes.}
\end{figure}
\end{document}